\documentclass{emulateapj}

\newcommand{\noun}[1]{\textsc{#1}}
\providecommand{\tabularnewline}{\\}

\begin{document}

\title{NGC~5011C: an overlooked dwarf galaxy in the Centaurus A group
\footnote{Based on observations made with ESO Telescopes at the La Silla Observatory}}

\author{Ivo Saviane}

\affil{European Southern Observatory, A. de Cordova 3107, Santiago, Chile}

\email{isaviane@eso.org}

\and{}

\author{Helmut Jerjen}

\affil{Research School of Astronomy and Astrophysics, The Australian National
University, Mt Stromlo Observatory, Cotter Road, Weston ACT 2611,
Australia}

\email{jerjen@mso.anu.edu.au}

\date{Received xxx; accepted xxx}

\begin{abstract}
A critical study of the properties of groups of galaxies can be done
only once a complete census of group members is available. Despite
extensive surveys, even nearby groups can lead to surprises. Indeed
we report the discovery of a previously unnoticed member of the Centaurus~A
Group, NGC~5011C. While the galaxy is a well known stellar system
listed with a NGC number its true identity remained hidden because
of coordinate confusion and wrong redshifts in the literature. NGC~5011C
attracted our attention since, at a putative distance of $45.3$~Mpc,
it would be a peculiar object having a very low surface brightness
typical of a dwarf galaxy, and at the same time having the size of
an early-type spiral or S0 galaxy. To confirm or reject this peculiarity,
our immediate objective was to have the first reliable measurement
of its recession velocity. The observations were carried out with
EFOSC2 at the 3.6m ESO telescope, and the spectra were obtained with
the instrument in long-slit mode. The redshifts of both NGC~5011C
and its neighbor NGC~5011B were computed by cross-correlating their
spectra with that of a radial velocity standard star. We found that
NGC~5011C has indeed a low redshift of $v_{\odot}=647\pm96\,{\rm km\, sec}^{-1}$
and thus is a nearby dwarf galaxy rather than a member of the distant
Centaurus cluster as believed for the past 23 years. Rough distance
estimates based on photometric parameters also favor this scenario.
As a byproduct of our study we update the redshift for NGC~5011B
at $v_{\odot}=3227\pm50\,{\rm km\, sec}^{-1}$. Applying population
synthesis techniques, we find that NGC~5011B has a luminosity-weighted
age of $4\pm1$~Gyr and a solar metallicity, and that the luminosity-weighted
age and metallicity of NGC~5011C are $0.9\pm0.1$~Gyr and $1/5$
solar. Finally we estimate a stellar mass of NGC~5011C comparable
to that of dwarf spheroidal galaxies in the Local Group. 
\end{abstract}


\section{Introduction}

During a CCD imaging campaign with the 40-inch telescope at Siding
Spring Observatory carried out by one of us (HJ) in 1999, the galaxy
NGC~5011C only 7.7 arcmin away from the target galaxy ESO~269-G070
was falling into the $20\times20$\,sq\,arcmin field-of-view and
attracted the attention because of its morphological appearance resembling
closely that of a nearby dwarf elliptical galaxy. Most strikingly
the stellar system exhibits the featureless light distribution with
the characteristic low central surface brightness. However, a search
through the literature for more details about this galaxy suggested
a different picture. Based on its position in the sky and published
redshift measurements of over 3000\,km\,s$^{-1}$, NGC~5011C should
be a member of the distant Centaurus cluster. If true, the immediate
question arises what kind of galaxy this is with a physical size larger
than that of the neighboring Centaurus cluster member, the S0 galaxy
NGC~5011B, but with the stellar surface density of a low luminous
dwarf galaxy (see Fig.~1)?

The galaxy NGC~5011C (ESO~269-G068) is located in the general direction
of the Cen\,A group and the Centaurus cluster. The nearby Cen\,A
group has a mean distance from the Milky Way of 3.7~Mpc (Jerjen,
Freeman \& Binggeli \citeyear{jerjen_etal00b}; Karachentsev et al.~\citeyear{karachentsev_etal06})
and a mean heliocentric velocity $\langle v_{\odot}\rangle=551$\,km\,s$^{-1}$
(C\^{o}t\'{e} et al.~\citeyear{cotge_etal97}). On the other hand,
the Centaurus cluster is $\approx12$ times further away at 45.3\,Mpc
(Mieske, Hilker \& Infante \citeyear{mieske_etal05}). Dickens, Currie
\& Lucey (\citeyear{dickens_etal86}) reported a redshift of 3195\,km\,s$^{-1}$
for NGC~5011C (galaxy \#531), consistent with the 3200\,km\,s$^{-1}$
from Fairall (\citeyear{fairall_83}), which immediately implies an
association with the Cen\,30 component of the Centaurus cluster that
has a mean velocity of $\approx3041$\,km\,s$^{-1}$ (Lucey, Currie
\& Dickens \citeyear{lucey_etal86}). From its location, NGC~5011C
seems to form a physical pair with the other Centaurus cluster member,
the S0 galaxy NGC~5011B (ESO~269-G067), which is only 1.1\,arcmin
to the North (Fig.~1) and has a very similar redshift of 3294\,km\,s$^{-1}$
as reported by the same authors (galaxy \#530). However, despite the
small projected distance of only 14\,kpc there is no obvious optical
sign of interaction between the two.

More confusion was added to the picture by the results from Dressler
(\citeyear{dressler_91}). The galaxy catalog listed a redshift only
for NGC~5011C (3293\,km\,s$^{-1}$, galaxy \#456) while no measurement
was reported for NGC5011B. This appeared somewhat strange as the central
surface brightness of NGC5011B is approximately four magnitudes higher
than that of NGC5011C making the former galaxy a more logical target
for a redshift measurement.

The present paper reports on our spectroscopic and optical follow
up of the galaxy pair to shed more light on this confusing situation
around the two galaxies NGC5011B and C. In the next section we describe
our 3.6m observations and data reduction, and the age and metallicity
of the stellar populations dominating the galaxies light are determined
in \S3. The new recession velocities for the two galaxies are presented
in \S4 followed by a photometric analysis with a rough distance estimate
for NGC5011C in \S5 and a discussion in \S6.

\section{Spectroscopy}

The observations were carried out with the ESO~3.6m telescope at
La Silla, during an EFOSC2 (Buzzoni et al. \citeyear{buzzoni_etal84})
technical night, with the purpose of testing all science and calibration
templates, and to estimate the photometric zero points and the grism
efficiencies. The {}``rotate to slit'' template was tested by putting
the nuclei of NGC~5011B and C into the same 1\,arcsec slit, and
taking $3\times1200$\,sec spectra. Before and after the galaxies'
observation, spectrophotometric standards were observed, and immediately
after the end of the last spectrum, $3\times1$\,sec spectra of a
radial velocity standard were also taken. Although several grisms
were tested during the night, for this particular set of data grism
\#11 was used, since it gives a good wavelength coverage and its intermediate
dispersion allows to obtain a good S/N for faint targets in relatively
short times. An observing log is listed in Table~\ref{tabcap:Journal-of-the}.
The radial velocity standard is HD125184 (from Udry et al.~\citeyear{udry_etal99}),
and the rest of the objects in the table are spectrophotometric standards
from Hamuy et al.~(\citeyear{hamuy_etal92,hamuy_etal94}).

All reductions were carried out with the ESO-Midas data processing
system (e.g. Warmels \citeyear{warmels91}), and in particular within
the \noun{long} context, which provides commands to perform the standard
reduction steps of long-slit spectroscopy. A series of scripts were
written to facilitate as much as possible the batch reduction of several
spectra, and they are publicly available in the EFOSC2 instrument
pages. We first prepared the system to reduce the night spectra, using
the afternoon calibrations: the frames were rotated to have the wavelength
increasing along the CCD rows, a median bias frame was created and
subtracted from the rest of the frames, and a normalized flat-field
was also generated. To do this, five flat-fields taken with slit $1\farcs0$
were median combined, and then a two-dimensional polynomial was fitted
to the resulting image: the polynomial had degree 6 along the dispersion
axis, and it was constant along the spatial direction. This artificial
image was subtracted from the median flat, thus removing the lamp
continuum and leaving only the pixel-to-pixel variation. The wavelength
solution was established with a HeAr arc spectrum taken with the $1\farcs0$
slit. A database of wavelength solutions for each EFOSC2 grism is
also published in the instrument web pages, so by using the appropriate
guess session for grism \#11 it was straightforward to get the solution
for the current data set. After searching the master arc for emission
lines, and after defining a two-dimensional grid of points distributed
on the arc ridge lines, the grid was fitted with a two-dimensional
polynomial, with degree 3 along the dispersion axis, and degree 2
along the spatial direction, which is enough to correct the curvature
of the arc lines. The r.m.s. error of the wavelength solution is $0.26$~\AA,
and the calibrated spectra cover the range 3383 to 7505 \AA\ with
a dispersion of 4.23 \AA/pix. The width of the arc lines at 5000
\AA\ is $\approx13.3$~\AA, so a velocity measurement based on
a single spectral line has a 1-sigma error of $\approx340\,{\rm km\, sec}^{-1}$.

During the night, after a new spectrum was taken, it was reduced using
the calibration data prepared above: it was rotated and trimmed, bias-subtracted
and flat-fielded, and the wavelength calibration was applied, which
transforms the bidimensional frames from the pixel $(x,y)$ space
to the $(\lambda,s)$ space, where $s$ is the spatial direction after
correction for the distortion introduced by the grism. The one-dimensional
spectra were extracted with a simple average over the rows covered
by the profile, and then multiplied by the number of rows. The extraction
window was defined by inspecting the spatial profile of the spectra,
and taking the rows were the spectrum was significantly brighter than
the background. Prior to the extraction, the sky background was subtracted
from each frame: two template sky spectra were defined in two windows
preceding and following the target spectrum, and the sky background
at the position of the target was defined by a linear interpolation
between these two spectra.

The spectrophotometric standard LTT~3218 was used to define the response
function of the grism: in this case the spectrum was extracted using
a very wide window and then compared to the spectrum published by
Hamuy et al.~(\citeyear{hamuy_etal92}). The spectra were flux-calibrated
by first normalizing by their exposure time and then dividing by the
response function. Prior to this, they were corrected for atmospheric
extinction using the average curve at La Silla. The fully reduced
and calibrated spectra of the two galaxies are shown in Fig.~\ref{figcap:The-flux-calibrated-spectra}.

\section{Ages and metallicities}

The spectral energy distribution (SED) of NGC~5011B is obviously
redder than that of NGC~5011C, but a more quantitative comparison
of these two spectra is not possible, since the slit was aligned along
the two galaxy centers, and not along the parallactic angle. To overcome
this problem, we started from the fact that NGC~5011B is classified
as an S0 galaxy (Lauberts \citeyear{lauberts_82}). Looking at Fig.~1
of Zaritsky et al. (\citeyear{zaritsky_etal95}) one can see that
the SEDs of this class of galaxies are rather uniform, so by selecting
a template spectrum and comparing it to our observed one, we can estimate
the degree of loss of blue flux induced by the differential atmospheric
refraction. Zaritsky et al. used the spectral atlas of Kennicutt (\citeyear{kennicutt92}),
which includes five S0 galaxies. We chose NGC~3245 since, compared
to the other templates, it has no emission lines, its SED is not rising
in the blue, it extends to redder wavelengths, and it has better spectral
resolution. By dividing the NGC~5011B spectrum by the template (redshift
corrected), a correction function was found, which is smoothly declining
from red to blue, reaching at $3400$~\AA\ $60\%$ of the value
at $7500$~\AA. In other words, $40\%$ of the blue flux was lost
during the observation. The trend of the correction is well fit by
a straight line, which was then used to compute the intrinsic SEDs
of the two galaxies. The corrected spectra are shown in Fig.~\ref{figcap:The-flux-calibrated-spectra}
together with the original ones, and after normalizing their fluxes
to unity at $6000$~\AA. The original spectra of NGC~5011B and
NGC~5011C have fluxes reaching a maximum of $1.5\times10^{-15}$~erg~cm$^{-2}$~sec$^{-1}$~\AA$^{-1}$
in the $R$ band and $0.06\times10^{-15}$~erg~cm$^{-2}$~sec$^{-1}$~\AA$^{-1}$
in the $B$ band, respectively. In the $R$ band, NGC~5011C is four
magnitudes fainter than NGC~5011B.

A first check of the correctness of our SEDs can be done by using
them to perform synthetic broad-band photometry in order to compute
colors, and then comparing the result to that obtained with our direct
imaging. To compute synthetic colors, a stellar A0I template was taken
from Jacoby et al. (\citeyear{jacoby_etal84}) library, and instrumental
magnitudes and color were computed by integrating it within the Johnson
$B$ and $V$ filter passbands from Bessell (\citeyear{bessell05}).
A zero-point correction was then computed, to calibrate the color
to the standard Johnson system; the color of an A0I star ($B-V=-0.01$)
was taken from Allen's Astrophysical Quantities (Cox \citeyear{allen}).
By repeating this procedure with the two galactic SEDs as input, we
found $B-V=0.95$ and $B-V=0.47$ for NGC~5011B and NGC~5011C, respectively.
The spectrum of NGC~5011C has been extracted in the central $5\farcs3$,
and in Fig.~\ref{fig:sbprofiles} one can see that $(B-V)\approx0.4\pm0.08$
at the center of the galaxy (see below). The synthetic color is thus
in good agreement with that obtained from broad-band surface photometry,
which means that our correction indeed yields a reliable SED.

\begin{figure}
\plotone{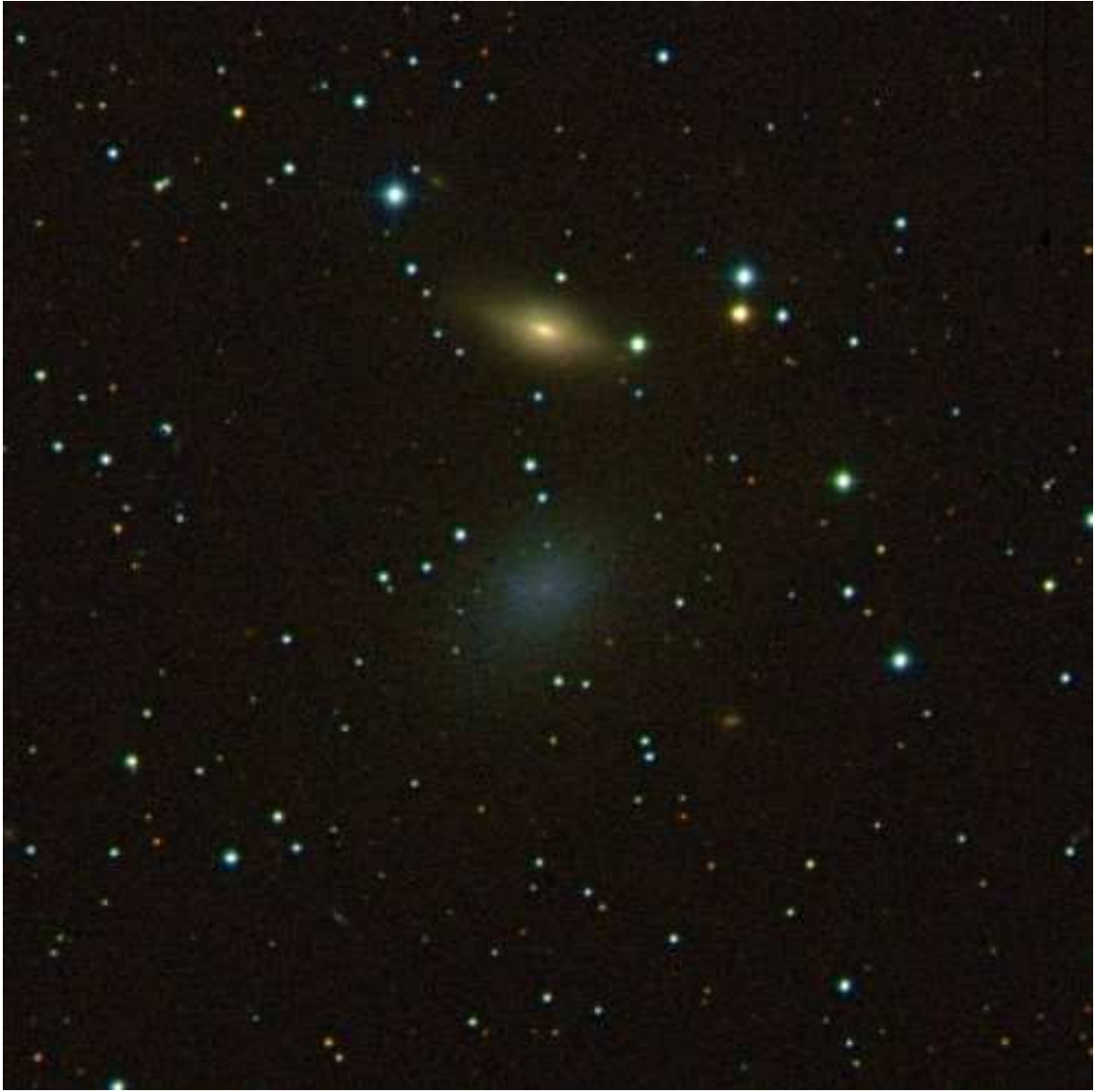}

\caption{EFOSC2 true-color image of NGC~5011B (top red galaxy) and NGC~5011C
(bottom blue galaxy). The composite was created from BVR frames and
the field of view is $\sim5\arcmin\times5\arcmin$. North is up, East
to the left. \label{figcap:EFOSC2-true-color-image}}
\end{figure}

\begin{figure*}

\plotone{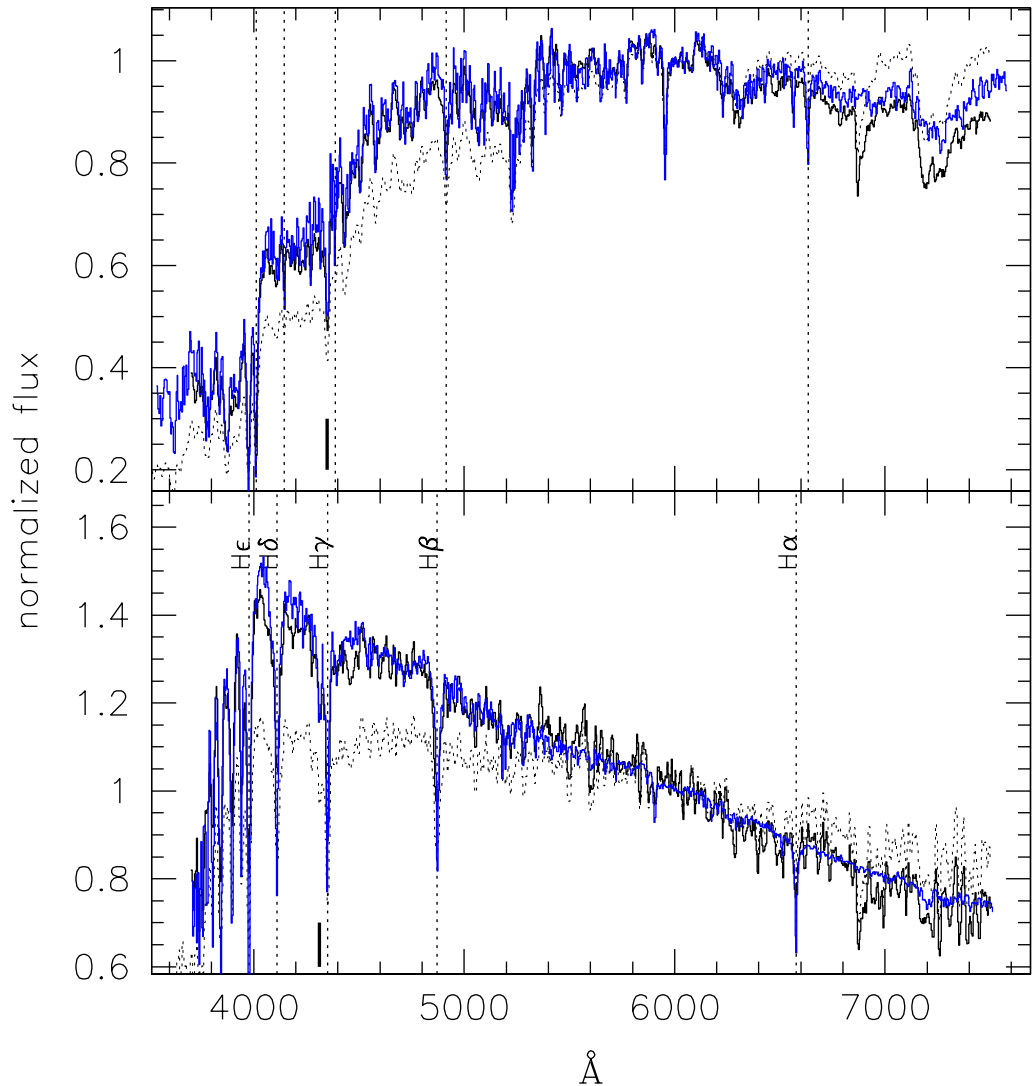}

\caption{The flux-calibrated, and normalized, spectra of NGC~5011B and NGC~5011C
are shown in the top and bottom panels, respectively. The dotted curves
show the original spectra, while the solid curves show the spectra
after correcting for the missing blue flux, and with their flux normalized
in the $[6000:6050]$\AA\ region. The spectrum of NGC~5011C has
been smoothed with a window of three pixels. The vertical dotted lines
show the location of the first five Balmer lines (H$\alpha$, $\beta$,
$\gamma$, $\delta$, $\epsilon$) red-shifted with the individual
velocities of the galaxies. The thick segment shows the (red-shifted)
location of the G band. NGC~5011C has evident Balmer lines and only
a hint of G band, while the G band and the H and K Ca lines at $\sim3950$~\AA\ are
clearly present in the spectrum of NGC~5011B. Moreover, only weak
H$\alpha$ and H$\beta$ can be seen. The blue curves are model single
stellar populations, computed for an age of $4$~Gyr and $0.9$~Gyr,
and a metallicity $Z=0.02$ and $Z=0.004$ for NGC~5011B and NGC~5011C,
respectively (see text for details).\label{figcap:The-flux-calibrated-spectra} }
\end{figure*}

After this consistency check, we can interpret the two SEDs using
population synthesis techniques. For that purpose we used GALAXEV
from Bruzual \& Charlot (\citeyear{bc03}; BC03), to estimate  the
luminosity-weighted age and metallicity. A grid of model SEDs were
generated, based on the Padova 1994 tracks (see BC03 for the references)
with an input Salpeter stellar mass function. Computations were done
for ages ranging from $0.1$~Gyr to $20$~Gyr, and for five of the
available metallicities $Z=0.0001$, $0.0004$, $0.004$, $0.008$,
and $0.02$. The age was increased in steps of $0.1$~Gyr up to $1$~Gyr,
and in steps of $1$~Gyr for older ages, and the spectra were degraded
to our instrumental resolution with software provided by BC03. The
best-fit was searched by minimizing the difference between model and
observed SEDs. In this way we found that the best solution is obtained
for an age of $0.9\pm0.1$~Gyr and $Z=0.004$ for NGC~5011C, and
for an age of $4\pm1$~Gyr and $Z=0.02$ for NGC~5011B. 
%
%
In the BC03 scale the metallicity of the Sun is $Z=0.02$, so the
two metallicities are $1/5$ solar and solar, respectively.
The good
match between the theoretical and observed SEDs can be checked in
Fig.~\ref{figcap:The-flux-calibrated-spectra}. Note that these results
only apply to the central regions of the two galaxies; for example
the NGC~5011C spectrum was extracted in the central $5\farcs3$,
which correspond to $95$~pc at the estimated distance
of the galaxy (see below). The age and metallicity thus represent
the central stellar population only, and since the rest of the galaxy
has redder colors presumably it hosts older populations, as is evident
from Fig.~\ref{fig:sbprofiles}. Indeed a kind of tiny, whiter, bulge-like
core can be seen in Fig.~\ref{figcap:EFOSC2-true-color-image} as
well.

The total $V$ magnitude computed in \S\ref{sec:Surface-Photometry}
allows us to estimate the mass of NGC~5011C as well. At a fiducial
distance of $3.7$~Mpc it converts into $M_{V}\sim-13.9$, and assuming
$M_{V}^{\odot}=4.7$, the total luminosity can be computed. Finally,
from Fig.~1 of BC03 one can estimate that, for an age of $1$~Gyr,
the mass-to-light ratio is $M/L_{V}\approx0.3$, so the mass of the
galaxy is $M\sim 8 \times 10^{6}M_{\odot}$. This value of the mass is only
valid in the single age approximation, but it's very likely that the
SFH of the galaxy is more extended, so realistically that mass represents
only a lower limit. But even with this caveat, the stellar mass is
in the regime of that of dwarf galaxies.

\section{Recession velocities of NGC 5011B and C}

\label{sec:Recession-velocities-of}

\begin{figure}[t]
\plotone{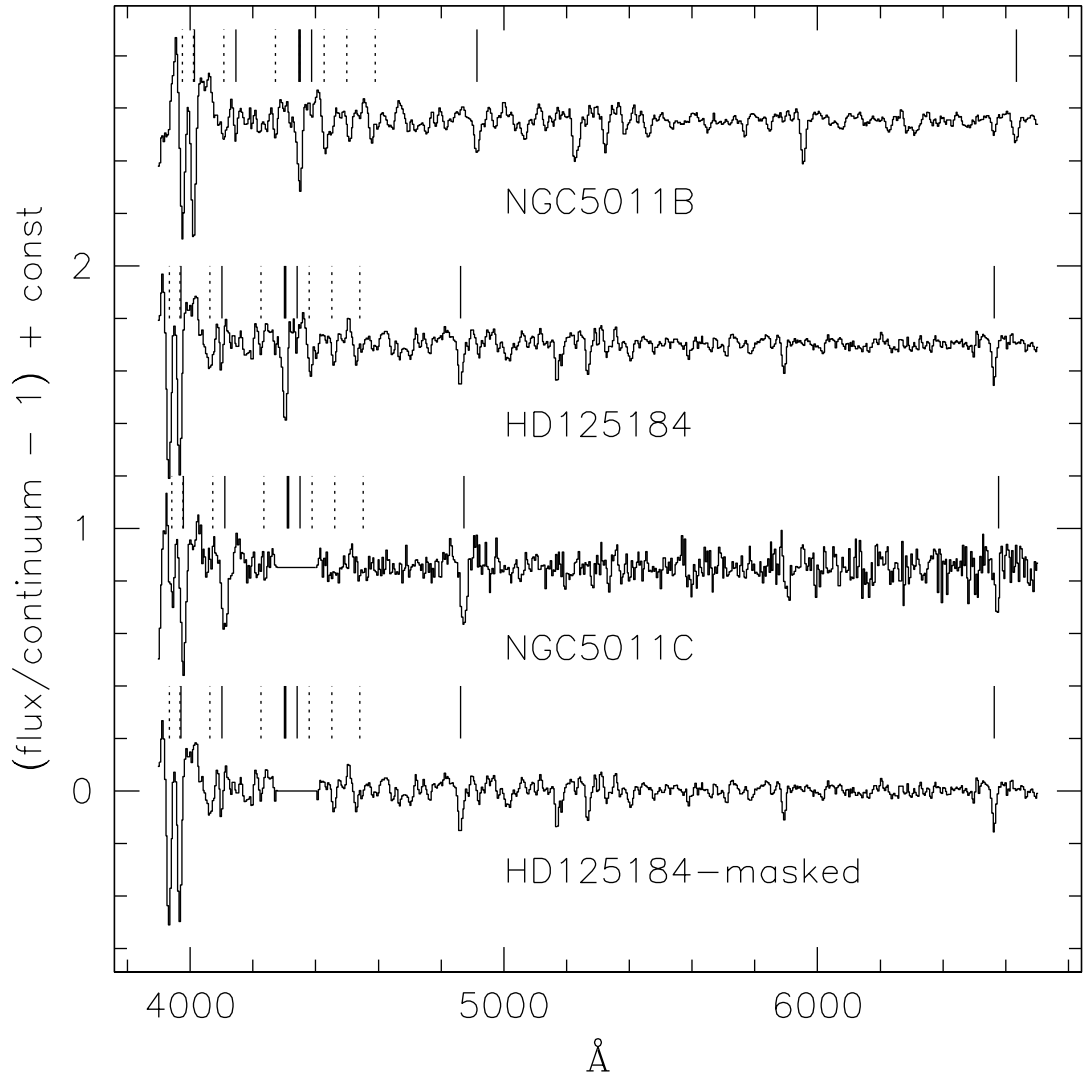}

\caption{The radial velocities were computed by cross-correlating the upper
two and lower two spectra shown in this figure. The upper two continuum-normalized
spectra are those of NGC~5011B and the RV standard; the lower two
spectra are those of NGC~5011C and the same standard, after masking
the H$\gamma$ and G band region. The vertical solid segments identify
Balmer lines, while the dotted ones identify features typical of G-type
stars. In particular the heavy segment marks the position of the G
band. For the two galaxies, the wavelengths of the lines have been
red-shifted using the individual recession velocities computed in
Sec.~\ref{sec:Recession-velocities-of}. \label{fig:The-upper-two}}
\end{figure}

A brief inspection of the prominent Balmer lines in the spectra (see
Fig.~\ref{figcap:The-flux-calibrated-spectra}) quickly reveals that
the two galaxies must have quite different redshifts in contrast to
what has been reported in the literature. An initial estimate of the
recession velocities can be obtained by measuring the central wavelength
of one of these lines. For example, the center of H$\beta$ is at
$4915$\AA\ in the NGC~5011B spectrum but at $4873$\AA\ in the
NGC~5011C spectrum (see also Fig.~\ref{fig:The-upper-two}). This
translates into a velocity of $3290\pm340\,{\rm km\, sec}^{-1}$ for
NGC~5011B while that of NGC~5011C would be $750\pm340\,{\rm km\, sec}^{-1}$,
where the error is the minimum uncertainty estimated above at $\sim5000$\AA.
To compute more precise velocities, we cross-correlated our spectra
with that of the radial velocity standard HD125184, thus obtaining
the relative velocity of each galaxy with respect to that of the standard
star. If we call this relative velocity $\Delta v^{{\rm gal}}$, then
the true heliocentric velocity of the galaxy is\[
v_{\odot}^{{\rm gal}}=\Delta v^{{\rm gal}}+v_{\odot}^{{\rm std}}+\delta v_{\odot}^{{\rm gal}}-\delta v_{\odot}^{{\rm std}}\]
 where $\delta v_{\odot}$ is the heliocentric correction listed in
Table~\ref{tabcap:Journal-of-the}, and the heliocentric velocity
of the standard star is $v_{\odot}^{{\rm std}}=-12.40\pm0.05\,{\rm km\, sec}^{-1}$.
From the same table, one can see that $\delta v_{\odot}^{{\rm gal}}-\delta v_{\odot}^{{\rm std}}=-3.7\,{\rm km\, sec}^{-1}$,
where the individual heliocentric corrections have been computed using
\noun{iraf}'s task \noun{rvcor}. The cross-correlation of the spectra
has been done using \noun{iraf}'s task \noun{fxcor}. To improve the
signal, the spectra were first trimmed to $3900-6700$\AA\
and then continuum-normalized, as shown in Fig.~\ref{fig:The-upper-two}.
Additionally, in the case of NGC~5011C the region containing H$\gamma$
and the G band was masked out before cross-correlating the two spectra:
the two features are located close to each other at our spectral resolution
and as H$\gamma$ is basically absent in the RV standard spectrum,
the algorithm would overestimate the velocity as it attempts to correlate
H$\gamma$ in the galaxy spectrum with the G band in the standard
star spectrum. Masking the two spectra accordingly removed the confusion
leading to a more realistic velocity estimate with a reduced measurement
error. At the end of this analysis we get $\Delta v^{{\rm NGC5011B}}=3243\pm50\,{\rm km\, sec}^{-1}$
and $\Delta v^{{\rm NGC5011C}}=663\pm96\,{\rm km\, sec}^{-1}$, so
the heliocentric velocities are $v_{\odot}=3227\pm50\,{\rm km\, sec}^{-1}$
and $v_{\odot}=647\pm96\,{\rm km\, sec}^{-1}$ for NGC~5011B and
NGC~5011C, respectively.

\section{Surface Photometry \label{sec:Surface-Photometry}}

To gain insight into NGC5011C's colors and photometric parameters,
shallow broad-band B, V, and R frames of 90s, 60s, and 30s integration
time, respectively, were acquired with the EFOSC2 instrument during
the same observing night (see Table~\ref{tabcap:Journal-of-the}).
A montage image is shown in Fig.~\ref{figcap:EFOSC2-true-color-image}.
The two galaxies have similar angular size, but they are of entirely
different morphological type. NGC~5011B is a classical lenticular
(S0) galaxy seen almost edge-on, while NGC~5011C appears as a low
surface brightness, early-type system with a uniform color and a hint
of a tiny nucleus or central star concentration.

Continuous MeteoMonitor measurements of the atmospheric conditions
at La Silla observatory showed that the extinction RMS was less than
0.02~magnitudes indicating that the sky transparency was good. From
9--11 measurements spread in time the photometric zero points for
the three passbands were calculated: ZP(B)$=26.06\pm0.07$, ZP(V)$=26.12\pm0.03$,
and ZP(R)$=26.19\pm0.04$. The pre-reduction of the raw CCD frames
was carried out within MIDAS in the usual manner. After the bias subtraction,
pixels that do not respond linearly to flux were replaced by a median
of the surrounding area, using the latest EFOSC2 bad pixel mask, and
then each science frame was flattened using high signal-to-noise sky
flats. All reduced frames were flat to less than 0.08\%. The sky level
was determined from star-free regions well away from the galaxies.
Then foreground stars and background galaxies were removed from the
images to allow uncontaminated photometric measurements using procedures
written within the IRAF package. Stars in the vicinity of NGC5011C
were replaced by small patches of empty sky. The area underneath the
neighbor galaxy NGC5011B was restored using a rectangular sub-image
from the opposite side of the galaxy. The center of NGC5011C was given
by the intensity-weighted centroid.

The galaxy flux was integrated using circular apertures on the clean
$BVR$ images to produce growth curves as a function of the geometric
radius $r=(ab)^{1/2}$ (where $a$ and $b$ are the major and minor
axis). The total magnitude $m_{T}$ was determined by extrapolation
to the asymptotic intensity, and the effective radius $r_{{\rm eff}}$
was defined as the value at which the intensity reached half of $m_{T}$.
Within $r_{{\rm eff}}$ the mean surface brightness $\langle\mu\rangle_{{\rm eff}}$
was also computed. These photometric parameters are listed in Table
\ref{photdata}. Surface brightness profiles in the three bands and
extinction-corrected colour gradients (B--R)$_{0}$, (V--R)$_{0}$,
and (B--V)$_{0}$ are shown in Figure \ref{fig:sbprofiles}. Thereby
we used the extinction values A$_{B}=0.503$, A$_{V}=0.386$, and
A$_{R}=0.311$ from Schlegel, Finkbeiner \& Davis (\citeyear{schlegel_etal98}).
Additionally, S\'{e}rsic parameters were measured by fitting a S\'{e}rsic
model (S\'{e}rsic \citeyear{sersic_68}): \[
\mu(r)=\mu_{0}+1.086\,(r/r_{0})^{n}\]
 to the observed profiles. These model parameters are also listed
in Table~\ref{photdata} and the corresponding model curves superimposed
onto the surface brightness profiles in Fig.~\ref{fig:sbprofiles}.
The quoted uncertainties include formal errors owing to the sky background
removal, photometric uncertainty and the fit to the surface brightness
profile. The total magnitudes are in good agreement with the values
published in the ESO-Uppsala catalogue (Lauberts \& Valentijn \citeyear{luberts_valentijn89}).

Although the current imaging data is not deep enough to measure an
accurate distance for NGC5011C by means of the surface brightness
fluctuations method (e.g.~Jerjen et al.~\citeyear{jerjen_etal01})
we can make a guess at the distance interval in which NGC5011C is
likely to be found with the help of the two independent quantities
$\langle\mu\rangle_{{\rm eff}}$ and $n$ if we accept the galaxy
to be of early-type morphology.

(A) The $\langle\mu\rangle_{{\rm eff}}-m_{B}$ relation for early-type
dwarf galaxies in the Virgo cluster (Binggeli \& Cameron \citeyear{binggeli_cameron91})
predicts an apparent B magnitude of $15.5\pm1.5$\,mag for a galaxy
like NGC5011C with an extinction-corrected mean effective B surface
brightness of $\langle\mu\rangle_{{\rm eff}}^{0}=22.94$ mag arcsec$^{-2}$
at the Virgo distance. Employing the latest cluster distance modulus
of $(m-M)_{{\rm Virgo}}=31.05$\,mag (Jerjen, Binggeli \& Barazza
\citeyear{jerjen_etal04}, Mei et al.~\citeyear{mei_etal06}) this
translates into an absolute magnitude of $M_{B}=-15.55\pm1.5$\,mag.
Combining that number with the apparent B magnitude in Table \ref{photdata}
we infer a distance modulus of $(m-M)=m_{B}-A_{B}-M_{B}=29.85\pm1.5$\,mag
or a distance range of 4.7--18.6\,Mpc.

(B) As an alternative, we can use the S\'{e}rsic parameter $n$ that
controls the overall shape of a galaxy to estimate the distance interval.
Equation (6) from Binggeli \& Jerjen (\citeyear{binggeli_jerjen98})
that describes the $\log(n)-m_{B}$ relation for early-type dwarfs
in the Virgo cluster predicts an apparent B magnitude of $m_{B}=16.29\pm0.92$
mag for NGC5011C at the Virgo distance. Using the same cluster distance
modulus as before yields an absolute magnitude of $-14.76\pm0.92$
mag, a galaxy distance modulus of $29.06\pm0.92$\,mag, and a distance
range of 4.3--9.9\,Mpc.

The main conclusion we can draw from these estimates based on photometric
properties is that the derived distance intervals are consistent with
NGC5011C being a member of the Cen\,A Group at 3.7\,Mpc (Jerjen,
Freeman \& Binggeli 2000; Karachentsev \citeyear{karachentsev05})
but incompatible with the Centaurus cluster distance of 45.3\,Mpc
(Mieske, Hilker \& Infante \citeyear{mieske_etal05}).

\section{Discussion and Conclusions}

The deep spectroscopy and shallow imaging with the 3.6m ESO telescope
has confirmed a new member of the nearby Cen\,A Group whose true
identity remained hidden because of coordinate confusion and wrong
redshifts in the literature since 23 years. The galaxy's morphology
provided the deciding clue. NGC~5011C is only 2.25 degrees to the
West of the dominant group galaxy NGC~5128 (Centaurus~A). Adopting
a mean group distance of 3.7\,Mpc this angular separation translates
into 150\,kpc. The two galaxies have also comparable velocities with
$v_{\odot}=647\pm96\,{\rm km\, sec}^{-1}$ for NGC~5011C and $v_{\odot}=547\pm5\,{\rm km\, sec}^{-1}$
for NGC~5128. The sky distribution of all known group members in
the vicinity of NGC~5128 is shown in Fig.~\ref{figcap:distribution}.
Heliocentric velocities and distances taken from the literature (Jerjen,
Binggeli \& Freeman \citeyear{jerjen_etal00a}; Karachentsev \citeyear{karachentsev05};
Karachentsev et al.~\citeyear{karachentsev_etal06}) are given in
parentheses.

NGC~5011C is well within the velocity distribution of the Cen~A
group, and it is definitely not compatible with the velocity distribution
of the Centaurus cluster. The Cen~30 component of the cluster has
a velocity dispersion of $\sigma_{v}=586$~km~sec$^{-1}$, so NGC~5011C's
velocity is $4\,\sigma$ lower than the cluster's average. Although
very high peculiar velocities can be observed in galaxy clusters,
this is not the case for the Centaurus cluster, and all its members
have velocities greater than $1500$~km~sec$^{-1}$ (see Fig.~3
in Lucey et al. \citeyear{lucey_etal86}).

At 3.7\,Mpc, NGC~5011C has an absolute blue magnitude of $M_{B}=-13.54$\,mag
that makes it the 8th brightest member of the group comparable in
luminosity to ESO~269-066 and ESO~384-016. In terms of photometric
properties such as the overall (B--R)$_{0}$ color and the S\'{e}rsic
parameters the galaxy resembles closely the intermediate type (dS0/Im)
dwarf ESO~384-016 (see Table 2 of Jerjen, Binggeli \& Freeman \citeyear{jerjen_etal00a}).

NGC~5011C has a bluer global appearance ($\langle$B--R$\rangle_{0}\approx1.1$\,mag)
than NGC~5011B but unlike for instance the blue LSB galaxy UGC~5889
(Vallenari et al.~\citeyear{vallenari_etal05}) it shows no sign
of on-going star formation neither in the 2D image (e.g.~young star
clusters, \ion{H}{2} regions) nor in our 1D-spectrum (emission lines).
Our estimated age of $0.9$~Gyr for the central stellar population,
and the presumably older disk, is compatible with this scenario, since
much younger stars would be needed to ionize any existing gas.

NGC~5011C is neither a compact blue dwarf nor does it exhibit an
irregular stellar distribution or spiral arms. Optically, the galaxy
has smooth, symmetric isophotes suggesting a classification as dwarf
elliptical or intermediate (dE/Im) type galaxy.

A look at the 21\,cm spectrum from the {H}\textsc{i} Parkes All-Sky
Survey (Barnes et al.~\citeyear{barnes_etal01}) observed in the
direction of NGC~5011C gives the impression of some emission at the
optical velocity (see Fig.~\ref{figcap:HIPASS}). However, the periodic
baseline structure in the velocity range 150\,km\,s$^{-1}<cz<1500$\,km\,s$^{-1}$
makes it impossible to draw a final conclusion on the possible presence
or absence of \ion{H}{1} in NGC~5011C. At the tentative distance
of 3.7\,Mpc a non-detection corresponds to an upper limit of $M_{{\rm HI}}\sim1.4\times10^{7}\, M_{\odot}$
which is still a factor of 2.5 higher than the {H}\textsc{i} mass
of $(6.0\pm0.5)\times10^{6}\, M_{\odot}$ detected in ESO~384-016
(Beaulieu et al.~\citeyear{beaulieu_etal06}). For a most recent
list of \ion{H}{1} detections and upper limits in Cen A group dwarf
galaxies see Bouchard et al. (\citeyear{bouchard_etal06}). 

To obtain a better {H}\textsc{i} mass estimate and to measure an
accurate distance from surface brightness fluctuations or the tip
magnitude of the red giant branch will be part of a follow-up study
of NGC~5011C.

\acknowledgements{We acknowledge helpful discussions with Dominique Naef and Linda
Schmidtobreick, and we thank the anonymous referee for comments that
helped to improve the results of this work. We made use of the HIPASS
public data release. The Parkes telescope is part of the Australia
Telescope which is funded by the Commonwealth of Australia for operation
as a National Facility managed by CSIRO. This paper has been completed
thanks to a stay of I.S. at Mt Stromlo Observatory, supported both by
ESO (through a Director General grant) and the Australia National Unversity
(grant ARC DP0343156).}

\clearpage

\clearpage 

\clearpage %

\clearpage

\begin{figure}
\plotone{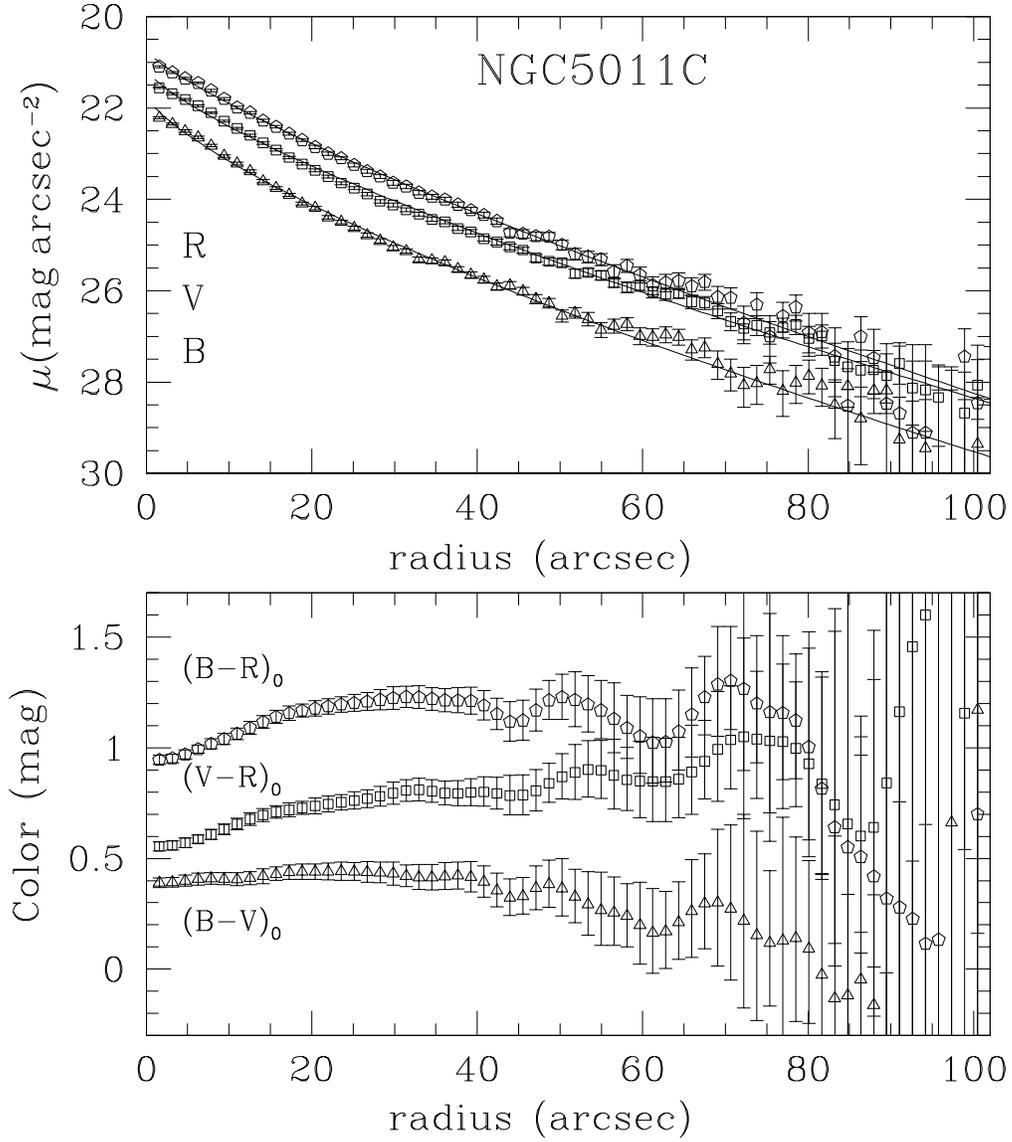}

\caption{(Top panel): BVR surface brightness profiles for NGC5011C as a function
of geometric radius $r=(ab)^{1/2}$, where $a$ and $b$ are the major
and minor axis. The best-fitting analytical Sersic profiles are superimposed
as black lines. (Bottom panel): extinction-corrected color gradients
smoothed with a third-order polynomial. \label{fig:sbprofiles}}
\end{figure}

\clearpage %
\begin{figure}
\plotone{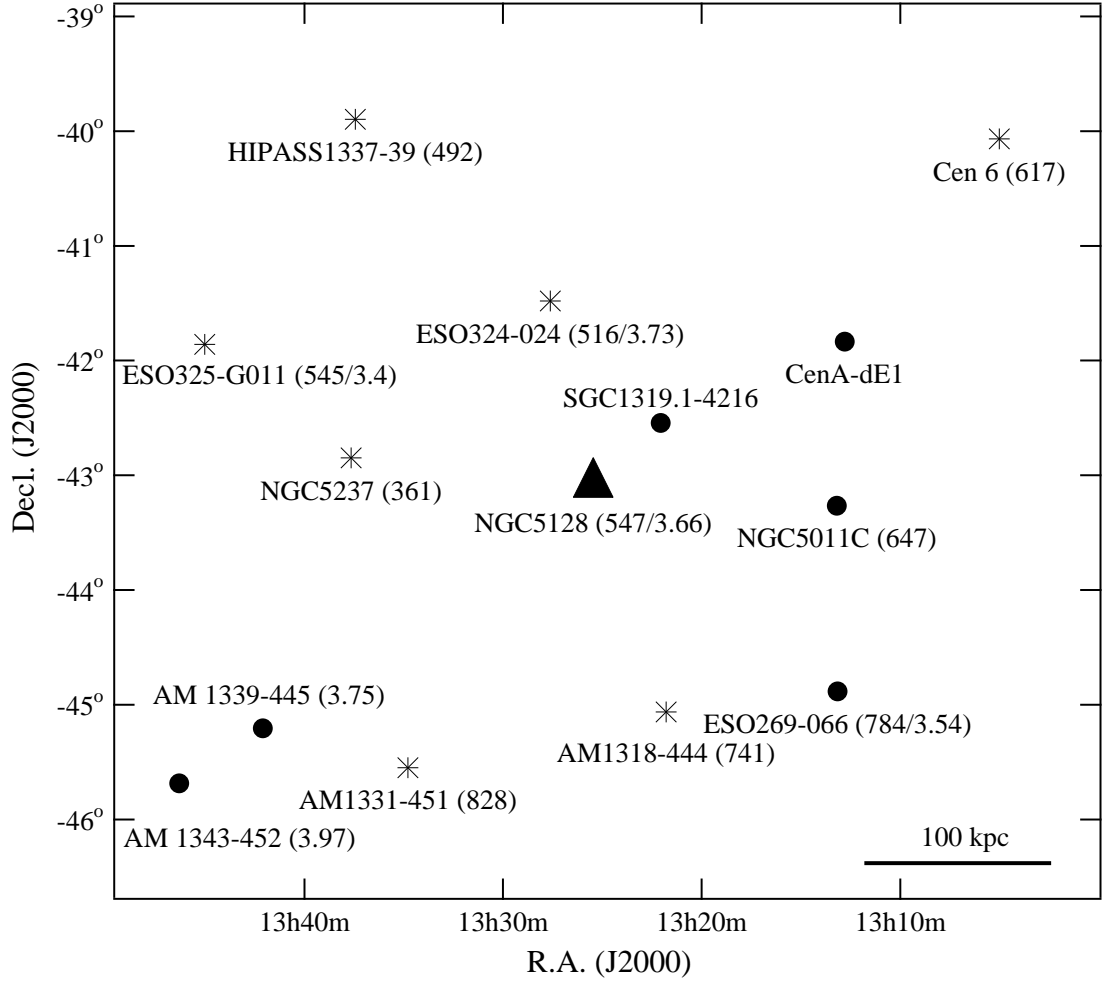}

\caption{The galaxy distribution around NGC~5128 (triangle). Early-type galaxies
are indicated as filled circles and late-type galaxies as stars. Heliocentric
velocities (in km\,sec$^{-1}$) and distances (in Mpc) are given
in parenthesis if available from the literature. \label{figcap:distribution}}
\end{figure}

\clearpage %
\begin{figure*}[th]
 \plotone{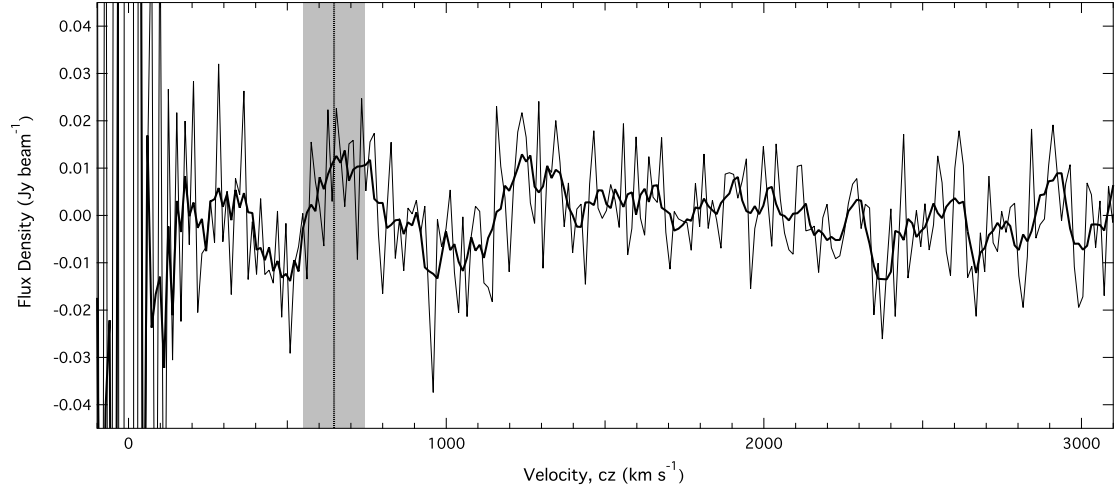}

\caption{The 21\,cm spectrum in the direction of NGC~5011C obtained from
the HIPASS public data release - v1.2 May 13, 2000 (south). The vertical
line and the grey band indicate the velocity and the 1$\sigma$ velocity
uncertainty as derived from our redshift measurement. Superimposed
is a 5-passes boxcar smoothed version of the spectrum (thick line)
that suggests a possible signal from {H}\textsc{i} emission at the
optical position. However the periodic baseline structure in the velocity
range 150\,km\,s$^{-1}<v<1500$\,km\,s$^{-1}$ makes it difficult
to draw a final conclusion. }

\label{figcap:HIPASS} 
\end{figure*}

\clearpage

\begin{table}

\caption{Journal of the observations. \label{tabcap:Journal-of-the}}


\begin{centering}\begin{tabular}{ccccccccc}
&
&
&
&
&
&
&
&
\tabularnewline
\hline
\hline 
{\scriptsize UT start }&
{\scriptsize Target }&
{\scriptsize R.A.(J2000) }&
{\scriptsize DEC(J2000) }&
{\scriptsize t$_{{\rm exp}}$(sec) }&
{\scriptsize slit }&
{\scriptsize Filter/Grism}&
{\scriptsize Airmass}&
{\scriptsize $\delta v_{{\rm \odot}}$(km~sec$^{-1}$) }\tabularnewline
\hline 
{\scriptsize 2006-03-02T06:59:26}&
{\scriptsize NGC5011B/C }&
{\scriptsize 13 13 11.95 }&
{\scriptsize $-$43 15 56.0 }&
{\scriptsize 30 }&
{\scriptsize -- }&
{\scriptsize R }&
{\scriptsize 1.033 }&
{\scriptsize -- }\tabularnewline
{\scriptsize 2006-03-02T07:00:32}&
{\scriptsize NGC5011B/C }&
{\scriptsize 13 13 11.95 }&
{\scriptsize $-$43 15 56.0 }&
{\scriptsize 60}&
{\scriptsize -- }&
{\scriptsize V }&
{\scriptsize 1.032 }&
{\scriptsize -- }\tabularnewline
{\scriptsize 2006-03-02T07:02:11 }&
{\scriptsize NGC5011B/C }&
{\scriptsize 13 13 11.95 }&
{\scriptsize $-$43 15 56.0 }&
{\scriptsize 90}&
{\scriptsize -- }&
{\scriptsize B }&
{\scriptsize 1.032 }&
{\scriptsize -- }\tabularnewline
\hline 
{\scriptsize 2006-03-02T00:06:12 }&
{\scriptsize LTT2415 }&
{\scriptsize 05 56 24.30 }&
{\scriptsize $-$27 51 28.8 }&
{\scriptsize 40 }&
{\scriptsize 5$\arcsec$ }&
{\scriptsize \#11 }&
{\scriptsize 1.001 }&
{\scriptsize $-$17.85 }\tabularnewline
{\scriptsize 2006-03-02T00:26:40 }&
{\scriptsize LTT3218 }&
{\scriptsize 08 41 34.10 }&
{\scriptsize $-$32 57 00.1 }&
{\scriptsize 40 }&
{\scriptsize 5$\arcsec$ }&
{\scriptsize \#11 }&
{\scriptsize 1.153 }&
{\scriptsize $-$4.80 }\tabularnewline
{\scriptsize 2006-03-02T01:18:05 }&
{\scriptsize LTT3218 }&
{\scriptsize 08 41 34.10 }&
{\scriptsize $-$32 57 00.1 }&
{\scriptsize 40 }&
{\scriptsize 5$\arcsec$ }&
{\scriptsize \#11 }&
{\scriptsize 1.058 }&
{\scriptsize $-$4.88 }\tabularnewline
{\scriptsize 2006-03-02T03:28:29 }&
{\scriptsize LTT3218 }&
{\scriptsize 08 41 34.10 }&
{\scriptsize $-$32 57 00.1 }&
{\scriptsize 40 }&
{\scriptsize 5$\arcsec$ }&
{\scriptsize \#11 }&
{\scriptsize 1.015 }&
{\scriptsize $-$5.10 }\tabularnewline
{\scriptsize 2006-03-02T04:58:15 }&
{\scriptsize LTT4816 }&
{\scriptsize 12 38 50.94 }&
{\scriptsize $-$49 47 58.8}&
{\scriptsize 120 }&
{\scriptsize 5$\arcsec$ }&
{\scriptsize \#11 }&
{\scriptsize 1.137 }&
{\scriptsize 17.92 }\tabularnewline
{\scriptsize 2006-03-02T05:16:36 }&
{\scriptsize NGC5011B/C }&
{\scriptsize 13 13 12.13 }&
{\scriptsize $-$43 14 47.3 }&
{\scriptsize 1200 }&
{\scriptsize 1$\arcsec$ }&
{\scriptsize \#11 }&
{\scriptsize 1.130 }&
{\scriptsize 20.73 }\tabularnewline
{\scriptsize 2006-03-02T05:36:36 }&
{\scriptsize NGC5011B/C }&
{\scriptsize 13 13 11.88 }&
{\scriptsize $-$43 15 55.7 }&
{\scriptsize 1200 }&
{\scriptsize 1$\arcsec$ }&
{\scriptsize \#11 }&
{\scriptsize 1.097 }&
{\scriptsize 20.70 }\tabularnewline
{\scriptsize 2006-03-02T05:56:36 }&
{\scriptsize NGC5011B/C }&
{\scriptsize 13 13 12.00 }&
{\scriptsize $-$43 15 21.5 }&
{\scriptsize 1200 }&
{\scriptsize 1$\arcsec$ }&
{\scriptsize \#11 }&
{\scriptsize 1.072 }&
{\scriptsize 20.68 }\tabularnewline
{\scriptsize 2006-03-02T06:28:37 }&
{\scriptsize HD125184 }&
{\scriptsize 14 18 00.73 }&
{\scriptsize $-$07 32 32.6 }&
{\scriptsize 1 }&
{\scriptsize 1$\arcsec$ }&
{\scriptsize \#11 }&
{\scriptsize 1.210 }&
{\scriptsize 24.42 }\tabularnewline
{\scriptsize 2006-03-02T06:29:37 }&
{\scriptsize HD125184 }&
{\scriptsize 14 18 00.73 }&
{\scriptsize $-$07 32 32.6 }&
{\scriptsize 1 }&
{\scriptsize 1$\arcsec$ }&
{\scriptsize \#11 }&
{\scriptsize 1.208 }&
{\scriptsize 24.42 }\tabularnewline
{\scriptsize 2006-03-02T06:30:37 }&
{\scriptsize HD125184 }&
{\scriptsize 14 18 00.73 }&
{\scriptsize $-$07 32 32.6 }&
{\scriptsize 1 }&
{\scriptsize 1$\arcsec$ }&
{\scriptsize \#11 }&
{\scriptsize 1.206 }&
{\scriptsize 24.43 }\tabularnewline
{\scriptsize 2006-03-02T09:07:31 }&
{\scriptsize LTT6248 }&
{\scriptsize 15 39 00.02 }&
{\scriptsize $-$28 35 33.1 }&
{\scriptsize 40 }&
{\scriptsize 5$\arcsec$ }&
{\scriptsize \#11 }&
{\scriptsize 1.009 }&
{\scriptsize 29.15 }\tabularnewline
\hline
\end{tabular}
\par\end{centering}

\tablecomments{ For the target {}``NGC5011B/C'', the coordinates of NGC~5011B,
of NGC~5011C, and of a point in-between the two galaxies, are listed
in the three rows. Likewise the heliocentric velocity correction has
been computed for the three positions and UTs, in order to show that
an average correction can be safely adopted.}
\end{table}

\clearpage %
\begin{table}[htdp]

\caption{Photometric Parameters}

\begin{centering}\begin{tabular}{lccc}
&
$B$ &
$V$ &
$R$ \tabularnewline
\hline 
$m_{T}$ (mag) &
$14.80\pm0.08$ &
$13.95\pm0.07$ &
$13.50\pm0.06$ \tabularnewline
$r_{{\rm eff}}$ (arcsec) &
$21.3\pm0.7$ &
$23.3\pm0.5$ &
$22.3\pm0.2$ \tabularnewline
$\langle\mu\rangle_{{\rm eff}}$ (mag as$^{-2}$) &
$23.44\pm0.08$ &
$22.77\pm0.07$ &
$22.23\pm0.06$ \tabularnewline
\hline 
$\mu_{0}$ (mag as$^{-2}$) &
$21.71\pm0.21$ &
$21.15\pm0.14$ &
$20.74\pm0.16$ \tabularnewline
$r_{0}$ (arcsec) &
$6.7\pm1.6$ &
$8.3\pm1.2$ &
$9.2\pm1.3$ \tabularnewline
$n$ (S\'{e}rsic) &
$0.73\pm0.07$ &
$0.76\pm0.05$ &
$0.81\pm0.05$ \tabularnewline
\hline
\end{tabular}\par\end{centering}

\label{photdata} 
\end{table}

\end{document}